\documentclass[prd,superscriptaddress,altaffilletter,amssymb,showpacs,nofootinbib,twocolumn]{revtex4}
\usepackage[dvips]{graphicx} 
\usepackage[usenames]{color}
\usepackage[latin1]{inputenc}
\usepackage[english]{babel}
\newcommand{\be}{\begin{equation}}
\newcommand{\ee}{\end{equation}}
\newcommand{\bea}{\begin{eqnarray}}
\newcommand{\eea}{\end{eqnarray}}

\newcommand{\ben}{\begin{eqnarray}}
\newcommand{\een}{\end{eqnarray}}
\newcommand{\n}{\label}
\newcommand{\no}{\noindent}
\newcommand{\ga}{\gamma}

\newcommand{\La}{\Lambda}

\begin{document}
\title{Bridging geometries and potentials in   DBI cosmologies}
\author{Luis P. Chimento}
\email{chimento@df.uba.ar}
\affiliation{Departamento de F\'isica,
Universidad de Buenos Aires, 1428 Buenos Aires, Argentina}
\author{Ruth Lazkoz}\email{ruth.lazkoz@ehu.es}
\affiliation{Fisika Teorikoa, Zientzia eta Teknologia Fakultatea, Euskal Herriko Unibertsitatea, 644 Posta Kutxatila, 48080 Bilbao, Spain}

\date{\today}

\begin{abstract}
We investigate the link between the warp function and the potential in DBI cosmologies in connection with the possibility they represent power-law solutions. 
A prescription is given to take advantage of the known result that given a warp factor 
there is always a choice of potential resulting in a constant ratio 
between pressure and energy density. The method is illustrated with examples with interesting models for either
the warp factor or the potential. We complete this investigation by giving a recipe to exploit symmetries in order to generate new solutions from
existing ones; this method can be applied, for instance, to the power-law cosmologies obtained using our prescription. 
\end{abstract}
\maketitle

\section{introduction}

Consistent flavours of string theory allowing for open strings include 
a class of extended objects called D-branes to  which the end points of 
the open strings are attached. Those extended physical objects are 
typically non-perturbative features of string theory, and have 
therefore played a substantial role in the development of the theory 
(see \cite{Quevedo:2002fc} and references therein) 


The idea that the inflaton might be an excitation of a D-brane in the 
form of an opening string has captivated quite a few researchers (see \cite{Dvali:1998pa,Leigh:1989jq,Kachru:2003sx,Cline:2006hu,Kallosh:2007ig,HenryTye:2006uv,Silverstein:2003hf,Alishahiha:2004eh,Chen:2004hua,Chen:2005ad,
Kecskemeti:2006cg,Lidsey:2006ia,Baumann:2006cd,Spalinski:2007qy,Huang:2007hh,Bean:2007eh,Gmeiner:2007uw,Spalinski:2007dv,Spalinski:2007kt}
 and references therein for regular papers and \cite{Peiris:2007gz} for a review). The 
inflaton could be a mode describing the
position of a D3-brane  wandering (radially) in a ten-dimensional space--time with 
a warped metric.
 The interest of this interpretation of inflation is not just that it 
would brush it with
an aristochratic patina, it is a more pragmatic one, as in these 
stringy scenarios inflation
happens to occur without such strict requirements as in the standard 
weekly coupled slow roll inflation model. Specifically, inflation can 
proceed with much steeper potentials.


The motion of the brane seems to admit an effective good description in 
terms of a Dirac-Born-Infeld action coupled to gravity \cite{Alishahiha:2004eh}, and it results in a scalar field theory with non-
canonical kinetic terms.
The usual assumption is that the metric on the brane is a flat FRW one, and it is customary to take
advantage of the usual perfect fluid formulation of the corresponding Einstein equations. The modifications with 
respect to canonical scalar field models appear in the form of factors in front of the square of the 
gradient of this scalar field, which in turn depend on both the scalar field and its gradient also. This factor
is related to the speed of the wandering brane and investigations using inflationary parameters had been carried out to estimate 
how this non-canonicalities in the kinetic terms of the model
affect the cosmological observables  \cite{Silverstein:2003hf,Alishahiha:2004eh,Chen:2004hua,Chen:2005ad,Kecskemeti:2006cg,Lidsey:2006ia,Baumann:2006cd,Spalinski:2007kt,Huang:2007hh,Bean:2007eh,Gmeiner:2007uw}.

In this context of FRW DBI cosmologies there has been a considerable interest in possible realizations of 
power-law inflation, given the preeminence such evolutions have gathered 
over the years due to  their role 
as asymptotic equilibrium states \cite{wain}. 

An exceedingly well-known fact regarding power-law solutions is that, in 
terms of the customary perfect fluid interpretation (which applies also 
to DBI FRW cosmologies) power-law cosmologies correspond to a constant 
ratio between pressure and energy density. In these scenarios there are 
two degrees of freedom available which could be chosen so as to result 
is power-law cosmological models \cite{Spalinski:2007dv} that for a given warp factor 
there is always a choice of potential resulting in a constant ratio 
between pressure and energy density:  one of these free functions is the 
warp factor of the metric, which is denoted as $f(\phi)$, whereas the 
other is the inflaton potential $V(\phi)$. In this paper we provide an algorithm to exploit that correspondence in order to generate DBI solutions.
The novelty and  relevance of our method is that, as far as we are 
concerned, no exact (genuinely) DBI solutions exist in the literature, not even power-law ones. 
Of course, if one's aim is to enforce power-law solutions corresponding 
to a particular choice of either the warp factor or the potential, then 
some degree of compromise will be needed and approximations will perhaps be 
required \cite{Silverstein:2003hf,Alishahiha:2004eh,Spalinski:2007kt,Spalinski:2007qy,Spalinski:2007dv}, this was
indeed the route taken in precursor works. We advocate, however, that progress in the construction of 
power-law DBI cosmologies can be made along the path of exact formulations as 
well if the problem is posed in a different fashion and one rather seeks for 
a particular asymptotic form of the warp factor and potential rather than demanding it corresponds to
all regimes.

Specifically, we present neat expressions leading to a constructive recipe for DBI power-law cosmologies
which allows us for instance to show that the almost ubiquitous 
exponential potential can indeed accommodate that sort of solutions, provided one chooses an adequate 
warp factor. After analyzing this case we move on and conside others which under appropriate choices of
parameters and in asymptotic regimes lead to for instance the AdS throat warp factor \cite{Randall:1999vf}, or to  a
generalization of the inverse power-law potential which was earlier studied in connection with tachyon cosmologies \cite{Garousi:2004ph}.

In the course of the discussion we briefly comment on the conditions for the warp function $f$ to be positive (the contrary case is not consistent
from the string theory point of view, but it is admissible in the field theory spirit).
Finally, we also discuss about the possibility of generating further solutions from existing ones under the use of symmetries, and how
phantom DBI cosmologies could fit into this picture.

\section{DBI setting}
As mentioned in the Introduction, string theory provides a theoretical framework in which inflation can proceed without having to meet the tight requirement of the potential being rather flat as in conventional inflation models. According to this proposal
cosmic accelerated expansion (or inflation) can be the manifestation of the motion of an extended object 
(a D3-brane)  through a ten-dimensional space--time with a warped metric (see for e.g. \cite{Kecskemeti:2006cg}):
\begin{equation}
ds_{10}^2=\frac{1}{\sqrt{f(\phi)}}g_{\mu\nu}dx^{\mu}dx^{\nu}+\sqrt{f(\phi)}g_{mn}dy^mdy^n.
\end{equation}
Here $\phi$ is a single radial combination of the internal coordinates $y^m$.

Effectively, our scenario is that of a four-dimensional spatially flat FRW spacetime filled with a non-canonical scalar field. Using the customary perfect fluid interpretation we set
\begin{eqnarray}
 \rho=\frac{\gamma^2}{1+\gamma}\,\,\dot\phi^2+V(\phi)\label{rhodef},\\
p=\frac{\gamma}{1+\gamma}\,\,\dot\phi^2-V(\phi)\label{pdef},
\end{eqnarray}
with
\begin{equation}
\n{g}
\gamma=\frac{1}{\sqrt{1- f(\phi)\dot\phi^2}},
\end{equation}
and where, in principle, $f$ and $V$ are arbitrary functions.  Choosing the symbol $\gamma$ was not in origin accidental, but rather motivated by its analogy to the Lorentz factor of Special Relativity, as $\sqrt{f(\phi)}\dot\phi$ is interpreted as the proper velocity of the brane \cite{Silverstein:2003hf}. According to this the scalar field cannot 
roll down arbitrarily fast, because otherwise the speed treshold leading to nonanalytic behaviour of $\gamma$ would be trespassed. The slow roll limit of the model
corresponds to  $f(\phi)\dot\phi^2$, and in this nonrelativistic motion regime the familiar expressions for the energy the energy density and pressure of canonical scalar field
are recovered.

Assuming for the above fluid a barotropic equation of state  of the form $p=(\Gamma-1)\rho$, we get 
\be
\n{G}
\Gamma=-\frac{2\dot H}{3H^2}=\displaystyle\frac{\gamma\dot\phi^2}{\rho},
\ee
and the Einstein equations read
\begin{eqnarray}
&3 H^2=\displaystyle\frac{\ga\dot\phi^2}{\Gamma}, \n{00}&\\
&-2\dot H=\rho+p\equiv \gamma\dot\phi^2. \n{11}&
\end{eqnarray}
Even though our formulation may seem excesively concise,
the  information contained in the last two equations is basically all we need to elaborate our very general description of the realization of power-law inflation in this theoretical framework.

\section{Power-law solutions}

Making use of Eqs. (\ref{G})-(\ref{11}), the energy conservation equation can be written as
\be
\left(\frac{\gamma\dot\phi^2}{\Gamma}\right)^.+3H\gamma\dot\phi^2=0.
\ee
For $\Gamma=\Gamma_0$, upon integration we will obtain from Eq. (\ref{G}) the power-las scale factor $a\propto t^{2/3\Gamma_0}$. This means in this case (and this is actually what makes it appealing) the conservation equation gets readily integrated resulting in $
\gamma\dot\phi^2=c/a^{3\Gamma_0},$
where  $c$ is an integration constant. Such constant gets fixed upon replacement into the Friedmann equation, so we finally arrive at
\be
\n{a}
a=t^{2/3\Gamma_0}, \qquad \gamma\dot\phi^2=\frac{4}{3\Gamma_0\,t^2}.
\ee

The idea behind the procedure for reconstructing the warp factor $f$ and the potential  $V$ consistins in giving 
the scalar field as an invertible function of time, $\phi=\phi(t)$, so that finding $t=t(\phi)$ be possible. Obviously, one can also compute the derivative of the field with respect to time and then
reexpress it in terms of the field, $\dot\phi(t)=\dot\phi(t(\phi))$. This means that, if we are able to express both $f$ and $V$ in terms of time and the derivative of the field, then we will have solved the problem of finding   $f$ and $V$ as functions of $\phi$. Luckily, doing this is no hard task. From Eq. (\ref{g}) one can calculate the function $f$
\be
\n{f}
f=\frac{1}{\dot\phi^2}\left[1-\frac{9\Gamma_0^2}{16}\,t^4\dot\phi^4\right],\label{fsubs}
\ee
whereas, in the case $\Gamma=\Gamma_0$, and from Eqs. (\ref{rhodef}), (\ref{G})  and (\ref{f}),  
one can easily solve for the potential:
\be
\n{V}
V=\frac{4}{3\Gamma_0\,t^2}\left[\frac{1}{\Gamma_0}-\frac{4}{4+3\Gamma_0\,t^2\dot\phi^2}\right].\label{Vsubs}
\ee

Once again the known result that for any throat geometry there is a potential which
leads to power-law inflation for some range of parameters has become manifest. This may be viewed as a generalization to the also well-known facts that for $f=0$ (canonical scalar field models)
power-law inflation is possible if the potential is exponential \cite{Lucchin:1985} or the analogue result for the case $f=1$ (tachyon cosmologies) in which such kind of inflation
is obtained with inverse square potentials \cite{Bagla:2002yn,Feinstein:2002aj,Chimento:2003ta}.

The novelty here is we have overcome the apparent impossibility of finding analytical correspondences between the throat and the potential for power-law cases, our prescription allows progressing without needing to resort to the ultra-relativistic regime simplification. Once a given time dependence of $\phi$ is chosen the geometry of the warped metric can be recovered, the existence of inflation depends on the value of $\Gamma_0$
entering $f$ as a free parameter, and only values of $\Gamma_0$ giving a large enough $f$ will lead to inflationary behaviour. 

On the other hand, if one is to stick to an interpretation based on string
theory setting there is a restriction in the sign 
of the warp factor, as arbitrary choices of $\phi$ could lead to non strictly positive $f$; this would be problematic because the warped metric depends on the warp factor $f$ through its square root, so in the case of a negative warp factor the entries of the metric would become imaginary and
would need to consider some sort of analytic continuation, but we are not aware of results showing how this can be done.

There is however, the possibility to consider DBI actions from a field theory point of view, in which case a negative $f$ does not represent a problem. This is perfectly consistent and work along these lines has been carried out in \cite{Babichev:2006vx}, but one could argue the interest of providing a string theory interpretation of this negative $f$ models.

Having made these remarks, let us return to the main line we pursue here. In the next section we are going to exploit the above presented algorithm to elaborate on three examples. 

\subsection{Example 1}
Let us begin with the simple choice leading to the popular exponential (or  Liouville) potential. It stems from the choice
\begin{equation}
\phi=\frac{2}{A}\ln \vert{t}\vert.\label{fieldliou}
\end{equation}
with $A$ a constant.
Incidentally, it also leads to an  $f$  expressed in terms of a single exponential.  We find it convenient also to rearrange the parameters so that in the lines below the relation with the canonical scalar field will become clearer. From  the second bit of Eq. (\ref{a}) we find the relation $A^2=3\Gamma_0\ga_0$, with $\ga_0$ a constant. Then, combining  Eqs. (\ref{fieldliou}), (\ref{f}) and (\ref{V}) we get  $f$ and $V$; explicitly:
\be
f=\frac{A^2}{4}\left[1-\frac{1}{\ga_0^2}\right]e^{A\phi},
\ee
\be
V=\frac{4\ga_0^2}{A^2}\left[\frac{3}{A^2}-\frac{1}{1+\ga_0}\right]e^{-A\phi}.
\ee
Finally, the scale factor gets reexpressed in the following fashion:
\be
a=t^{2\ga_0/A^2}.
\ee
This way of formulating the solution is very interesting, as in the $\ga_0=1$ limit, that is, for $f=0$, the expressions just above go over to the simplified they get for the conventional canonical scalar field. 
This goes in consonance with the novel way of writing the energy density and pressure of the DBI fluid we have put forward. Note again how the new parametrizations (\ref{rhodef},\ref{pdef}) recover their conventional scalar field form for $\ga=\ga_0=1$. Nevertheless, it should be also noted that the $f=0$ does not admit either the string theory inspired
interpretation as it is the warped metric would blow up there, but the field theory interpretation remains of course valid.

\subsection{Example 2}
This example we consider now corresponds to
\be
\no \quad \phi=\phi_0 \,t^n, \quad  \phi_0=cons,
\ee
and it follows that
\be
\ga=\frac{4}{3\Gamma_0n^2\phi^2 }.
\ee
Once more, on using the latter one gets,
\be
f=\frac{\phi^{2/n-2}}{n^2\phi_0^{2/n}}\left[1-\frac{9\Gamma_0^2n^4}{16}\,\phi^4\right],
\ee
and
\be
V=\frac{4\phi_0^{2/n}}{3\Gamma_0\phi^{2/n}}\left[\frac{1}{\Gamma_0}-\frac{4}{4+3\Gamma_0n^2\phi^2}\right].
\ee
In the large $\phi$ regime $f$ goes like a negative power of $\phi$ for $-1<n<0$, and  like a positive power otherwise.
Actually, the large $\phi$ regime for this example is subject to the interpretational restrictions mentioned above as it characterized by $f <0$, and so the scheme is only valid in the field theory interpretation. In contrast, in the small $\phi$ regime $f$ goes like a negative power of $\phi$ for $n>1$, and like a positive power otherwise. Interestingly, the AdS throat often explored in the literature, i.e. $f\sim\phi^{-4}$ can be realized in our model either in a large $\phi$ regime 
if $n=-1/3$ or in a small $\phi$ regime if we rather consider $n=-1$.

The asymptotic behavior of the potential with  respect to the scalar field $\phi$ is simpler, as both in the large and small $\phi$ regimes we have $V\sim\phi^{-2/n}$.

\subsection{Example 3}
We now start off from the choice
\be\phi=\frac{2}{A}\ln \vert{B+t^n}\vert.
\ee 
with $A$ a constant.
Some straightforward steps involving use of 
Eqs. (\ref{fsubs},\ref{Vsubs}) allow obtaining $f$ and $V$ as functions of $\phi$.  
Our field choice leads to
\bea
f=\frac{A^2 e^{\scriptstyle {A \phi}}}{4 n^2\left(e^{\frac{A \phi}{2}}-B\right)^{2-\frac{2}{n}}}\left(1-\frac{9 n^4 \Gamma_0 ^2}{A^4e^{2 A \phi}} \left(e^{\frac{A
\phi}{2}}-B\right)^4 \right)\quad
\eea
and
\bea
V&=&\frac{4}{3\Gamma_0} \left(\frac{1}{\Gamma_0}-\frac{A^2 e^{A \phi}}{e^{A \phi} A^2+3
   \left(B-e^{\frac{A \phi}{2}}\right)^2 n^2 \Gamma_0 }\right)\times\quad \nonumber\\
&&\left(e^{\frac{A \phi}{2}}-B\right)^{-\frac{2}{n}}
\eea
If $\phi$ is non-negative, then strict positiveness of the warp factor is guaranteed if $3 n^2 \Gamma_0 \mbox{max}(1,B^2)<{A^2}$. 

\subsubsection{Case $B = 1$}
For this particular case it is not difficult to see  that for $1\gg t^n$ one approximately has
\bea
f\sim\phi^{{2}/{n}-2}.\\
V\sim \phi^{-{2}/{n}},
\eea 
which on the other hand are the same asymptotic expressions one has for the above example in its small field regime, and this is consistent as the   $\phi(t)$
expressions of the last two examples coincide at first order in the low field ($1\gg t^n$) regime.

If we  make in this regime the extra restriction that
$n\ge1$ in either Example 2 or 3, then form of the potential becomes approximately of inverse square type. It can, thus, be viewed as a generalization of the 
 $1/\phi^2$ for  k-essence and tachyon cosmologies.

\begin{figure}
\begin{tabular}{c}
\includegraphics[width=0.3\textwidth]{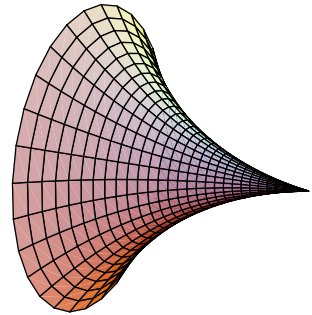}\quad\\
\includegraphics[width=0.3\textwidth]{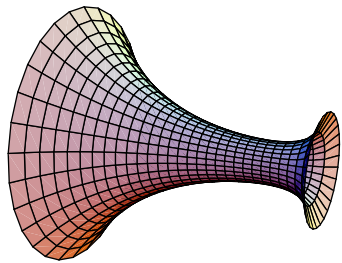}\quad
\end{tabular}
\caption{Warp factor for corresponding to Example 3 with $\Gamma_0=0.1$, $A=3$, and respectively $ n= 3$  and $B=0.2$ (upper figure) and 
$n=4$ and $B=0.7$ and $n=4$ (lower figure).}
\end{figure}

\subsubsection{Case $B\not = 1$}
For an arbitrary model the speed limit $\dot\phi^2\le f^{-1}(\phi)$ applies.
In the $B=1$ cases with $n>1$, this restriction makes it impossible for the scalar
field to reach the origin in finite time. This problem is absent, however, from their
 $B\not=1$ counterparts. This is advantageous in connection with reheating, as provided
 the potential has a minimum at the origin, then the field will be able to oscillate
 around it and and reheating will proceed. One can check that the necessary condition for the potentials discussed in this
 example to have a minimum at that location is
\begin{equation}
Gamma=\frac{2 A^2}{A \left(A\pm\sqrt{12 (B-1)^2 B n^3+A^2}\right)-6 (B-1)^2 n^2}
\end{equation}
  Of course, the power-law solutions one can obtain with the warp factor and potential presented in this example are not
 a representation of this oscillatory behaviour, we just want to bring about some nice properties of this model which
 make it interesting beyond its mere ability to acomodate power-law expansionary behaviour.

\begin{figure}
\begin{tabular}{c}
\includegraphics[width=0.3\textwidth]{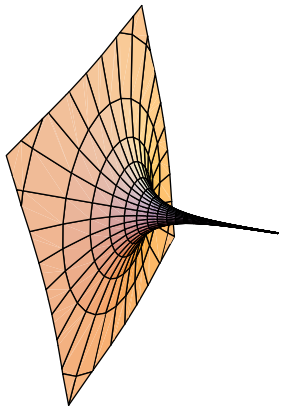}\quad
\end{tabular}
\caption{Warp factor corresponding to Example 3 with $\Gamma_0=0.9$, $A=3$, and respectively $ n=-1$  and $B=1$.}\end{figure}

\section{Duality}

In a series of papers \cite{Aguirregabiria:2003uh,Chimento:2003qy,Aguirregabiria:2004te,Chimento:2004br} the interest of form-invariance transformations as a method to obtain new exact solutions to the Einstein equations from
already existing ones has been shown. In the context of spatially flat perfect fluid FRW cosmologies, such transformations can be viewed
as a prescription relating the quantities $a$, $H$, $\rho$ and $p$ in a given initial scenario to quantities $\bar  a$, $ \bar H$, $\bar \rho$ and $\bar p$ corresponding to a new cosmological model.  As our investigation of power-law DBI cosmologies are concerned there is a class of form-invariance transformations which stands out, it is the one given by
$H\to \bar H=\eta H$, $ \rho\to\bar\rho=\eta^2\rho$, $p+\rho\to \bar p+\bar\rho=\eta (p+\rho)$ for an arbitrary real constant $\eta$. As will be shown below it allows
generating new power-law DBI cosmologies from existing ones, and if we particularize further to the case $\eta=-1$ we can consider it as a 
duality transformations, which provides a phantomization method (which is the way one could call the process of transforming a conventional cosmological model into a phantom one by performing a form-invariance transformation). Back to the general constant $\eta$ case, we would like to stress that our method provides the means to generate
inflationary power-law DBI cosmologies from non-inflationary ones.

Let us review the method from the very general perspective of a flat FRW spacetime filled with a perfect fluid. The Einstein equations read
with a perfect fluid
\begin{equation}
\label{00gen}
3H^{2}=\rho,
\end{equation} 
\be
\n{co}
\dot\rho+3H(\rho+p)=0,
\ee
where $\rho$ is the energy density, $p$ the pressure and $H=\dot a/a$ are invariant form under the
symmetry transformations
\bea
\label{tr}
&\bar\rho=\bar\rho(\rho)&\\
\label{th}
&\bar H=\left(\displaystyle\frac{\bar\rho}{\rho}\right)^{1/2}H&\\
\label{tp}
&\bar
p=-\bar\rho+
\left(\displaystyle\frac{\rho}{\bar\rho}\right)^{1/2}(\rho+p)\displaystyle\frac{d\bar\rho}{d\rho}&,
\eea
where $\bar\rho=\bar\rho(\rho)$ is an invertible function. Hence the FRW
equations for a perfect fluid have a form-invariance symmetry. The symmetry
transformations (\ref{tr})-(\ref{tp}) define a continuous Lie group which can be used to solve the FRW equations and get
accelerated expansion scenarios, as it will be seen in the following sections.

The symmetries we will exploit are of three different kinds: those with $|\eta|>1$ make the energy density of the universe bigger, those with  $\eta=\pm 1$ do not alter it, and those with $|\eta|<1$ do decrease. According with this taxonomy, and assuming the seed cosmology is an expanding one ($H>0$), the cosmology produced by the
transformation will also be an expanding one if  $\eta>0$, but alternatively be a contracting one for $\eta<0$. The special case $\eta=-1$ we dub dual symmetry, and it ensures the existence of a duality between a contracting universe filled with an ordinary fluid and an expanding universe driven by phantom energy. 

Now, it is interesting to investigate the transformation properties of the relevant physical parameters. For instance, the deceleration parameter
$q (t)=-H^{-2}(\ddot{a}/a)$, transforms as
\be
\n{tq}
\bar q+1=\left(\displaystyle\frac{\rho}{\bar\rho}\right)^{3/2}\displaystyle\frac{d\bar\rho}{d\rho}\,
(q+1),
\ee
 under the symmetry transformations (\ref{tr})-(\ref{tp}).

If we consider
perfect fluids with equations of state $p=\left(\Gamma-1\right)\rho$ and $\bar
p=\left(\bar\Gamma-1\right)\bar\rho$ respectively, then the barotropic
indices $\Gamma$ and $\bar\Gamma$ transform as
\be
\n{tg}
\bar\Gamma=\left(\displaystyle\frac{\rho}{\bar\rho}\right)^{3/2}\displaystyle\frac{d\bar\rho}{d\rho}\,
\Gamma,
\ee
under the symmetry transformations (\ref{tr})-(\ref{tp}). Besides, using
(\ref{tq}) and (\ref{tg}) we get a form invariant relation $(\bar
q+1)/\bar\Gamma=(q+1)/\Gamma$ between the deceleration parameter and the
barotropic index. As these results are readily applicable to any flat FRW power-law cosmological model (regardless of the theoretical
framework it fits in), they can be indeed exploited in connection with the DBI cosmologies considered in the previous sections. 

In general, inflationary solutions occur when $\ddot a>0$; this means that the expansion
is dominated by a gravitationally repulsive stress that violates the strong
energy condition, so that $\rho+3p<0$. Imposing this condition on (\ref{tq})
we obtain $d\bar\rho^{(-1/2)}/d\rho^{-1/2}<1/(q+1)$, which for a
non-accelerated cosmological model with $q\approx const >0$, gives
$\bar\rho>(q+1)^2\rho$. Such model, with $\bar q < 0$, is accelerated. This can be viewed understood in terms
of assisted inflation, as one achieves inflation by enhancing the energy density of the field. 
Recall we are considering
the possibility of having DBI cosmologies displaying  inflation, it should be clear that our transformations
can be used to generate new combination of geometry (throat) and potential with inflation starting perhaps
from others without that sort of expansionary behaviour, so the solution generation ability gets significantly enlarged.

So far we have progressed in a formal and rather general way, but it is convenient to provide further insight into the details of the transformation in the case we are concerned with. In order to avoid that the energy density and pressure turn into complex quantities we are going to impose the condition that the function $\ga$ remains invariant a given duality transformation; this implies the prodcut $f\dot\phi^2$ must transform into itself. It turns out that the form of corresponding Einstein equations do not change under the application of the following transformation:
 \begin{eqnarray}
&{\dot{\bar\phi}}^2=\eta{\dot\phi}^2&\n{1}\\
&\bar f=\eta f&\n{2}\\
&\bar V=(\eta^2-\eta)\rho+V=(\eta^2-\eta)\displaystyle\frac{\gamma^2}{1+\gamma}\dot\phi^2+\eta^2V& \n{3}
\end{eqnarray}
Consistently, it follows that the form-invariance of Eqs. (\ref{00gen}) and (\ref{11}) under the latter transformation requires $\Gamma=\Gamma/\eta$, so one goes from one power-law model into another.

A particular case of the latter transformation on which we are going to concentrate now is $\eta=-1$, i.e. the phantomization of the model, but before giving further details the remark is in order that the new solution of the dynamical equations corresponds to an imaginary field $\bar\phi=i\phi$ driven by a real potential  $\bar V=2\rho-V$ \cite{Chimento:2003qy,Aguirregabiria:2004te}

The phantomization process leads to a universe with  $\dot\rho>0$, which is equivalent to the violation of the weak energy condition $(\rho+p)<0$. 
It follows that two cases can be distinguished: a) $\rho$ has  $\rho\to\La$ asintote for $t\to\infty$ or $\rho$ grows unboundedly. In the first case the scale
factor ends up becoming that of the  de Sitter solution, that is $a\to \exp{\sqrt{\La/3}\,t}$. But in the other case, if we admit the asymptotic energy density  is $\rho\to \rho_0 a^\eta$ with $\eta>0$, then  the asymptotic solution to the Friedmann  equation is
\ben
\label{asin-}
a^-\to\left[\frac{2}{\eta\sqrt{\rho_0}\,(t_0-t)}\right]^{2/\eta}, \qquad t<t_0,\\
\label{asin+}
a^+\to\left[\frac{2}{\eta\sqrt{\rho_0}\,(t-t_0)}\right]^{2/n}, \qquad t>t_0.\\
\een
Clearly the expanding solution $a^-$ is defined for $t<t_0$ and displays a big rip at $t=t_0$ because the scale factor diverges at the finite time $t_0$ and a future singularity occurs. On the other hand the contracting solution $a^+$ begins at a past singularity at $t=t_0$. Summarizing, the solution $a^-$ arises by phantomization of the solution $1/a^-$ which ends with a big crunch at $t=t_0$. In terms of form-invariance transformations the phantomization is originated by the $1/a^-\to a^-$ duality existing between those two solutions to the Einstein equations.

In order to get a phantomization with a real potential and a real field \cite{Chimento:2004br} one must introduce a DBI$^{-}$ model with a sign reversal in the kinetic term entering the energy density and pressure (\ref{rhodef}) and (\ref{pdef}), so that
\begin{eqnarray}
 \rho^-=-\frac{\gamma^2}{1+\gamma}\,\,\dot\phi^2+V(\phi)\label{r-}\\
p^-=-\frac{\gamma}{1+\gamma}\,\,\dot\phi^2-V(\phi)\label{p-}
\end{eqnarray}
This allows enlarging the form invariance symmetry group, because now we can exchange not only solutions to the original equations (\ref{00})-(\ref{11}) among them, but also solutions to both equation sets among them.  In order to investigate this new enlarged symmetry group, let us rewrite Einstein equations in two form-invariant scenarios:
\ben
\n{01s}
3 H^2=s\frac{\ga\dot\phi^2}{\Gamma}, \qquad
-2\dot H=s\gamma\dot\phi^2,\\
\n{01b}
3\bar H^2=\bar s\frac{\bar\ga{\dot{\bar\phi}}^2}{\bar\Gamma}, \qquad
-2\dot{\bar H}=\bar s\bar\ga{\dot{\bar\phi}}^2,
\een
where $s=\pm 1$ y $\bar s=\pm 1$. In this case the transformations (\ref{1})-(\ref{3}) become
 \begin{eqnarray}
&{\dot{\bar\phi}}^2=-\displaystyle\frac{s}{\bar s} {\dot\phi}^2\n{1b}&\\
&\bar f=-\displaystyle\frac{\bar s}{s}f&\\
&\bar V=2\rho-V\n{3b}& 
\end{eqnarray}
This way it is possible now to achieve a phantomization with a real scalar field if we admit the existence of the theory DBI$^-$.

\section{Conclusions}
The so called DBI cosmologies are scalar field models depending with one additional functional degree of freedom as
compared to conventional scalar field configurations. This extra function has a geometrical meaning when DBI actions
are interpreted as describing the motion of a brane in a warped space time, the warp factor being indeed that additional
input required to specify the model. In this setup inflation is interpreted as the consequence of the motion of the brane in a background with extra dimensions, the compactification of which give rise to the scalar field's potential.

These scenarios have been profusely studied, and particular attention has been paid to the possibility they admit power-law solutions, as they seem to be favoured to play the role of equilibrium asymptotic states. In this paper
we have shown, that despite earlier claims pointing in the direction of finding such solutions in an exact way, 
completing such task is indeed possible. The method relies on providing a given parametrization of the scalar field
in terms of time, and then one can reconstruct the warp factor and the potential upon the sole requirement that the equation of state parameter be constant (in the fluid interpretation of the model). This powerful result is illustrated
by resorting to some examples with interesting asymtptotic limits coinciding with some models studied in the literature: the AdS throat, cut-off throats,  and the inverse square potential. 

Finally, in the last section, we show how to obtain new solutions  from existing ones using symmetries, specifically we
give transformations rules for the scalar field, warp factor and potential. This method has interesting applications as is offers the possibility of realizing assisted inflation in a DBI context, but also permits playing with the idea of phantom DBI cosmologies (to be generated from non phantom ones).

\section*{Acknowledgements}
We thank M. Spalinski for enlightening conversations.
L.P.C. is partially supported by the University of
Buenos Aires for partial support under project X224,
and the Consejo Nacional de Investigaciones Cient\'\i ficas
y T\'ecnicas under project 5169. R.L. is supported by the University of the Basque Country through research grant GIU06/37, and by the Spanish Ministry of Education and Culture through research grant FIS2004-01626.

\end{document}